\documentclass[reprint, amsmath,amssymb,
 aps,pra]{revtex4-2}
\usepackage{graphicx} % Required for inserting images
\usepackage[dvipsnames]{xcolor}
\usepackage{subcaption}
\usepackage{dcolumn}
\usepackage{bm}
\usepackage[hidelinks, colorlinks=true,linkcolor=MidnightBlue, citecolor=ForestGreen, urlcolor=Violet]{hyperref}
\usepackage{amsmath}
\usepackage{breqn}
\usepackage{amssymb}
\usepackage{physics}
\usepackage{multirow}
\usepackage{float}
\usepackage{cleveref}
\usepackage[normalem]{ulem}

\makeatletter
\let\cat@comma@active\@empty
\makeatother

\begin{document}

\setlength\parskip{0.2em}

\preprint{APS/123-QED}

\title{Vortex stability in interacting Bose-Einstein condensates}

\author{Ajay Srinivasan}\email{avsriniv@usc.edu}

\author{Aaron Wirthwein}\email{wirthwei@usc.edu}

\author{Stephan Haas}\email{shaas@usc.edu}
\affiliation{Department of Physics and Astronomy, University of Southern California, Los Angeles, CA 90007}

\date{\today}

\begin{abstract}
\noindent \textbf{Abstract:} We study the stability of vortices in a binary system of Bose-Einstein condensates, with their wave functions modeled by a set of coupled, time-dependent Gross-Pitaevskii equations. Beginning with an effective two-dimensional system, we identify miscible and immiscible regimes characterized by the inter- and intra-atomic interactions and the initial configuration of the system. We then consider a binary system of Bose-Einstein condensates placed in a rotating harmonic trap and study the single vortex state in this system. We derive an approximate form for the energy of a single vortex in the binary system and the critical angular velocity for the global stability of a vortex at the center of the trap. We also compute the metastability onset angular velocity for the local stability of a vortex at the center of the trap. Numerical solutions to the Gross-Pitaevskii equations support these expressions. These rotational results inform us of a novel subphase within the miscible regime of the binary condensate system. We thus demonstrate the non-trivial aspects of vortex stability in interacting binary Bose-Einstein condensates as a result of their non-linear interactions. 
\end{abstract}

\maketitle

\section{\label{sec:intro}Introduction}
The study of Bose-Einstein condensates (BECs) has significantly advanced our understanding of quantum mechanics on a macroscopic scale, revealing complex behaviors that are both theoretically interesting and experimentally observable. Among these behaviors, the formation and stability of vortices are particularly notable due to their complexity and relevance to quantum fluid dynamics. In previous work on the collision dynamics of one-dimensional Bose-Einstein condensates, we analyzed the interactions of two colliding BECs using coupled, time-dependent Gross-Pitaevskii equations \cite{wirthwein2022collision}. This study demonstrated how the interplay between inter- and intra-atomic interactions and initial configurations led to different dynamical regimes, from periodic transmission or reflection of condensates to more complex transient dynamics.

In this paper, we expand our investigation to a two-dimensional system to study vortex formation and stability in a binary system of interacting Bose-Einstein condensates. We begin by considering a binary system of vortex-free condensates, modeling their wave functions with coupled, time-dependent Gross-Pitaevskii equations. This approach allows us to identify distinct phases based on atomic interactions and the system's initial configurations. Subsequently, we analyze a binary BEC system confined in a rotating harmonic trap, focusing step-by-step on the formation and stability of a single vortex. We derive an approximate expression for the energy of a single vortex in a trapped system and calculate the critical angular velocities required for global and local stability of a vortex at the trap center. These findings are corroborated by numerical solutions of the Gross-Pitaevskii equations. Our study highlights the intricate behavior of vortices in binary Bose-Einstein condensates due to their nonlinear interactions, which can be observed under suitable experimental conditions.

This work builds on several previous studies addressing rotating trapped Bose-Einstein condensates \cite{Fetter_2008,bao_2006,Wieman_2000}. These earlier publications showed how rotating traps introduce discrete quantized vorticity, leading to vortex arrays and complex condensate behaviors under varying rotational speeds. They also developed efficient numerical methods for simulating these phenomena, emphasizing the conservation of angular momentum and the stability of vortex states. Our current study extends these insights by examining vortex formation and stability in \emph{interacting} binary BEC systems, providing a deeper understanding of these quantum systems and their potential applications.

\indent The paper is organized as follows. In Section \ref{sec:nonrot}, we discuss the formation of miscible and immiscible steady states in the absence of vortices. In Section \ref{sec:rot}, we analyze the stability of vortices in rotating condensates, exploring the effect of the trapping potential. The theoretical analysis in subsections \ref{sec:theory2:uni}, \ref{sec:theory2:trap}, and \ref{sec:theory2:rottrap} is complemented by numerical simulations in subsection \ref{sec:comp2}, indicating critical angular velocities which separate regimes of stable, metastable, and unstable vortices. In subsection \ref{sec:discussion2}, we discuss our results, particularly noting a novel \emph{dumbbell phase} within the miscible regime, informed by vortex stability. In \ref{sec:conclusion}, we summarize our results and outline limitations of our approach, as well as extensions of the framework to study broader classes of vortices and vortex lattices as in \cite{bao_2006, Wieman_2000, Coddington_2004, Fetter_2008, Fetter_2001, Kavoulakis_2000, Madison_2000}.

\section{\label{sec:nonrot} Two Interacting Vortex-Free Condensates}

Let us start by considering the formation of miscible and immiscible states in two interacting condensates composed of $N$ $^{87}\mathrm{Rb}$ atoms in the hyperfine spin states $|1\rangle = |1, -1\rangle$ and $|2 \rangle = |2,1\rangle$ \cite{wirthwein2022collision}.  We then write $a_{11}$ for the s-wave scattering length of a condensate composed of atoms in the $|1\rangle$ state, and $a_{22}$ for the s-wave scattering length of a condensate with atoms in the $|2\rangle$ state. The inter-atomic interaction between the condensates is parametrized by a scattering length $a_{12}$ (see Table I for experimentally measured values). We consider a harmonic potential trap,
\begin{equation}\label{theory1:1}
V(\mathbf{r}) = \frac{1}{2} m (\omega_{x}^2 x^2 + \omega_{y}^2 y^2 + \omega_{z}^2 z^2),
\end{equation}
where $m$ is the mass of Rubidium. The trap is designed with a sufficiently large $z$-confinement, so that the system is effectively two-dimensional in the $x-y$ plane. That is, $\omega_z$ is large enough so that the condensate is forced into the ground state along the $z$ direction but is free to explore excited states in the $x-y$ plane. Furthermore,  we initially focus on systems with vortex-free ground states. In Section \ref{sec:rot}, we derive criteria which the trap must satisfy to ensure that this is the case, but for now one may think of this as the study of a non-rotating trap or a sufficiently slowly spinning trap.

\begin{figure}[H]
    \includegraphics[width=0.45\textwidth]{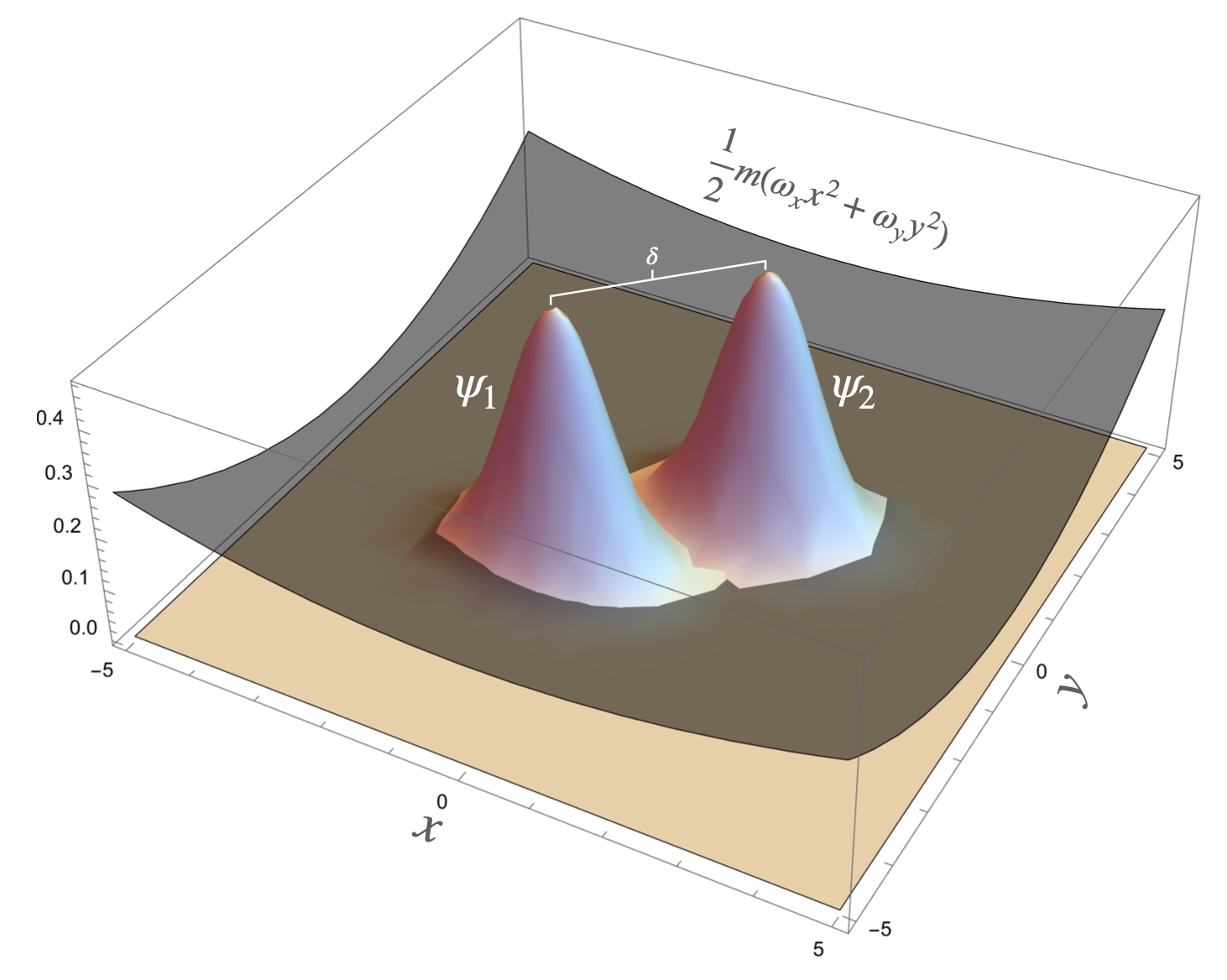}
    \caption{Initial configuration of two condensates with Gaussian profiles, denoted by $\psi_1$  and $\psi_2$, in a harmonic potential trap $V(x,y) = \frac{1}{2}m (\omega_x x^2 + \omega_y y^2)$. Here, $\delta$ is the distance between the centers of the condensates.}
    \label{fig:view2cartoon}
\end{figure}

\begin{table}[H]
    \centering
    \begin{tabular}{|p{4cm}|p{4cm}|}
\hline
\multicolumn{2}{|c|}{Model Parameters}\\
\hline
Quantity& Value or Expression\\
\hline
     $a_{11}$& 100.4$a_0$  \\
     $a_{22}$& 95.44$a_0$\\
     $a_{12}$& 98.006$a_0$\\
     $m$& 87 amu\\
     $\omega_{\perp}$& $2\pi \times 10$\\
     Atom number & 1000\\
\hline
\end{tabular}
    \caption{Parameters used throughout this work. Here $a_{jj}$ is the intra-atomic scattering length of the $j$-th condensate, $a_{12}$ is the inter-atomic scattering length, $m$ is the mass of a $^{87}\mathrm{Rb}$ atom, $\omega_{\perp}$ is the trap frequency, and $a_0$ is the Bohr radius. Since we are exploring the effect of interactions, the inter-atomic scattering strength is treated as a variable parameter.}
    \label{table:tab2}
\end{table}

The effectively two-dimensional dynamics of two vortex-free Rubidium condensates occupying different hyperfine spin states is modeled using the coupled Gross-Pitaevskii equations \cite{wirthwein2022collision} \cite{pitaevskii2003},
\begin{equation}\label{theory1:2}
\footnotesize i\hbar \frac{\partial \psi_1}{\partial t} = \left(-\frac{\hbar^2}{2m}\nabla^2 + V + Ng_{11}|\psi_1|^2 + Ng_{12}|\psi_2|^2\right)\psi_1,\end{equation}
\begin{equation}\label{theory1:3}
\footnotesize i\hbar \frac{\partial \psi_2}{\partial t} = \left(-\frac{\hbar^2}{2m}\nabla^2 + V + Ng_{22}|\psi_2|^2 + Ng_{12}|\psi_1|^2\right)\psi_2,
\end{equation}
where $N$ is the atom number, $\psi_1$ and $\psi_2$ are the 2D wave functions describing the $|1\rangle$ and $|2\rangle$ condensates respectively, and $g_{ij}$ are the renormalized interaction parameters introduced in \cite{wirthwein2022collision}.

We seek to understand how $g_{12}$, $\omega_x$, and $\omega_y$ affect the dynamics of the system late in the time-evolution, as a steady state is approached. Let us first identify the ground states of the system, using methods similar to those in \cite{wirthwein2022collision}. We begin with the energy functional,
\begin{dmath}\label{theory1:4}
    E = \iint \text{d}A \left(\sum_{i=1,2} \left(\frac{\hbar^2}{2m} |\nabla \psi_i|^2 + V|\psi_i|^2\\ + \frac{g_{ii}|\psi_j|^4}{2}\right) + g_{12}|\psi_1|^2|\psi_2|^2\right).
\end{dmath}

We then apply the variational method to investigate ground state properties of this system \cite{wirthwein2022collision}, considering the  Gaussian ansatz,
\begin{equation}\label{theory1:5}
    \psi_{j} = \frac{1}{\sqrt{\pi \sigma_{jx} \sigma_{jy}}} \exp\left(- \frac{(x-x_j)^2}{2\sigma_{jx}^2}\right) \exp\left(- \frac{(y-y_j)^2}{2\sigma_{jy}^2}\right).
\end{equation}
Here, for each $j=1,2$, $\sigma_{jx}$ and $\sigma_{jy}$ are variational parameters describing the width of the Gaussian in the $x$ and the $y$ directions respectively, while $x_j$ and $y_j$ are variational parameters describing the central location of the Gaussian. Fig. \ref{fig:view2cartoon} shows a visual depiction of our ansatzes in the harmonic trap, representing the initial configuration of our condensates which are modeled by two Gaussians with peak-separation $\delta$. 

    Inserting \eqref{theory1:5} into \eqref{theory1:4}, we find:
\begin{equation}\label{theory1:6}
\begin{split}
       E &= \frac{N\hbar^2}{4m} \left(\frac{1}{\sigma_{1x}^2} + \frac{1}{\sigma_{1y}^2} + \frac{1}{\sigma_{2x}^2}+\frac{1}{\sigma_{2y}^2}\right)\\
        &+ \frac{N}{4}m\omega_x^2 (\sigma_{1x}^2+2x_1^2 + \sigma_{2x}^2+2x_2^2)
        \\&+\frac{N}{4}m\omega_y^2(\sigma_{1y}^2 + 2y_1^2+\sigma_{2y}^2 + 2y_2^2)\\
        &+\frac{N^2}{4\pi} \left(\frac{g_{11}}{\sigma_{1x}\sigma_{1y}} + \frac{g_{22}}{\sigma_{2x}\sigma_{2y}}\right)\\
        &+ \frac{N^2 g_{12}\exp\left(-\left(\frac{\delta_x^2}{(\sigma_{1x}^2 + \sigma_{2x}^2)}+\frac{\delta_y^2}{(\sigma_{1y}^2+\sigma_{2y}^2)}\right)\right)}{\pi \sqrt{(\sigma_{1x}^2+\sigma_{2x}^2)(\sigma_{1y}^2+\sigma_{2y}^2)}}.
    \end{split}
\end{equation}
Here we have defined $\delta_x \equiv x_1 - x_2$ and $\delta_y \equiv y_1 - y_2$. The ground state conditions for the system can now be obtained using a variational analysis (\ref{app:a}) of Eq. \eqref{theory1:6}. For the remainder of this section, we assume that $g_{11}=g_{22} \equiv g$. Suppose at least one of $\delta_x$ and $\delta_y$ is non-zero (without loss of generality, let $\delta_x \neq 0$). Under this assumption, we find that $\sigma_{1x}=\sigma_{2x}=\sigma_x$ and $\sigma_{1y}=\sigma_{2y}=\sigma_y$. We thus obtain a balance relation between the widths of the condensates in the $x$ and $y$ directions, 
\begin{equation} \label{theory1:7} \sigma_x \omega_x = \sigma_y \omega_y.
\end{equation}

This analysis is restricted to the case of separation solely in the $x$-direction, i.e., $\delta_y=0$ and $\delta := \delta \geq 0$. In Appendix \ref{app:a}, we obtain Taylor approximations for $\sigma_x$, $\sigma_y$, and $\delta$ in the large $N$ limit where the kinetic energy contributions to the ground state conditions may be neglected. These approximations give
\begin{equation}\label{theory1:8}
\sigma_x^4 = \frac{N\omega_y}{m\pi \omega_x^3} \frac{g+g_{12}}{2},
\end{equation}
\begin{equation}\label{theory1:9}
\sigma_y^4 = \frac{N\omega_x}{m\pi \omega_y^3} \frac{g+g_{12}}{2},
\end{equation}
\begin{equation}\label{theory1:10}
\delta = \left(\frac{2N\omega_y}{m\pi \omega_x^3}(g+g_{12})\right)^{1/4} \sqrt{\log_{e} \left(\frac{g_{12}}{g}\right)}.
\end{equation}

\begin{figure}[h!]
    \centering
    \includegraphics[width=0.5\textwidth]{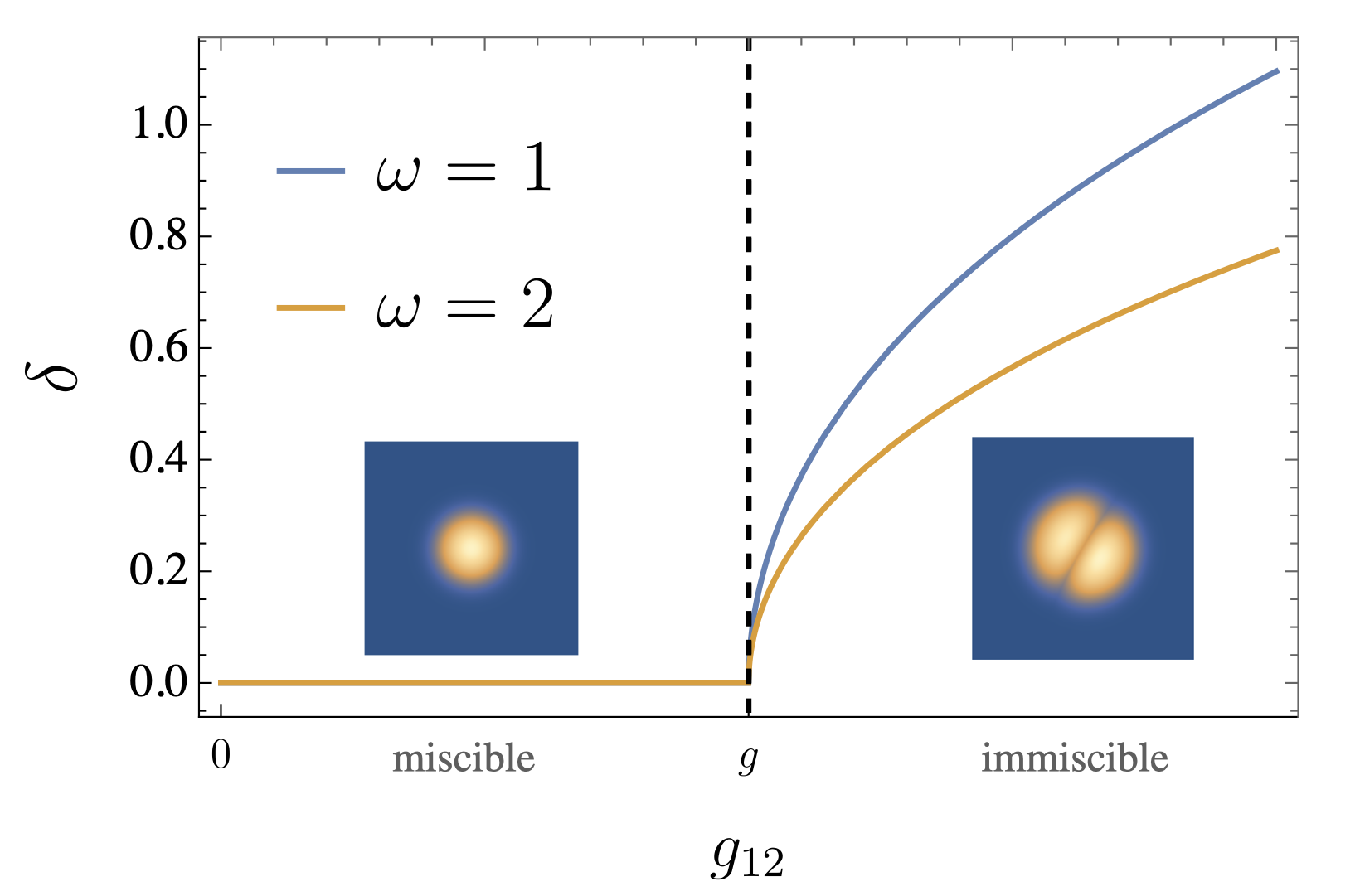}
    \caption{Dependence of the distance between the condensate centers,  $\delta$, on the inter-condensate repulsion strength $g_{12}$ and on the trap frequency $\omega$ (within the approximations considered in Section \ref{sec:nonrot}). Here, we consider the case where the trap frequency within the $x-y$ plane is $\omega = \omega_x = \omega_y$, and we set $2N/m\pi = 1$ and $g=1$. The  dotted vertical line demarcates the critical point at $g_{12}=g$. For $g_{12}<g$, the system is in a miscible phase shown by the inset on the left. For $g_{12}>g$, the system is in an immiscible regime illustrated by the inset on the right.}
    \label{fig:delta v w}
\end{figure}

Eq. \eqref{theory1:10} indicates a singularity at $g_{12}=g$, corresponding to a phase transition. For $g_{12}<g$, we note that $\delta_x=\delta_y=0$ must necessarily be true. For $g_{12}>g$, the ground state solutions are approximately described by the above analysis. In Fig. \ref{fig:delta v w}, we plot the relationship between the distance between condensate centers and the inter-condensate repulsion strength in Eq. 10 for two in-plane trapping frequencies, $\omega=\omega_x=\omega_y$. For $g_{12}<g$, the system resides in the miscible phase, depicted in the left inset. For $g_{12}>g$, the system transitions to the immiscible phase, shown in the right inset.

\section{\label{sec:rot} Two Interacting Rotating Condensates}
 
\subsection{\label{sec:theory2:uni} Uniform Case in Cylindrical Coordinates}
Next we address vortices in binary condensate systems placed in a harmonic trap, again considering two condensates composed of $N$ $^{87}$Rb atoms in the hyperfine spin states $|1 \rangle = |1, -1 \rangle$ and $|2 \rangle = |2, 1 \rangle$. Let $g_{11}$, $g_{22}$, and $g_{12}$ be the interaction parameters, as used in Section \ref{sec:nonrot}. We first focus on the single vortices in the uniform system ($V \rightarrow 0$) with $g_{12} < \sqrt{g_{11}g_{22}}$, i.e. the condensates are in the mixing regime.

    To proceed with the analysis of the GPEs, we switch to cylindrical coordinates, letting $\psi_j = f_j(\rho)e^{il\varphi}$, so that $|\psi_j| = f_j(\rho)$. Here $l$ is a non-negative integer describing the quantum of circulation of the vortex in question. The time-independent GPEs \cite{pethick2002bose} for the system are
\begin{equation}\label{theory2:27}
    \small -\frac{\hbar^2}{2m\rho}\frac{\mathrm{d}}{\mathrm{d} \rho}\left(\rho \frac{\mathrm{d} f_1}{\mathrm{d} \rho}\right)+\frac{\hbar^2 l^2 f_1}{2m\rho^2} + g_{11}f_1^3 + g_{12}f_2^2 f_1 = \mu_1 f_1,
\end{equation}
\begin{equation}\label{theory2:28}
    \small -\frac{\hbar^2}{2m\rho}\frac{\mathrm{d}}{\mathrm{d} \rho}\left(\rho \frac{\mathrm{d} f_2}{\mathrm{d} \rho}\right)+\frac{\hbar^2 l^2 f_2}{2m\rho^2} + g_{22}f_2^3 + g_{12}f_1^2 f_2 = \mu_2 f_2.
\end{equation}
Here, $\mu_j$ are the chemical potentials of each condensate. We now focus on vortices with $l=1$. At sufficiently large radial distances, the $1/\rho^2$ terms and radial derivatives may be neglected. Therefore, for large $\rho$ the GPEs reduce to a pair of coupled linear equations, which can be solved to yield asymptotic (in $\rho$) solutions,
\begin{equation}\label{theory2:29}
f_{1} \sim \frac{g_{22}\mu_1 - g_{12}\mu_2}{g_{11}g_{22}-g_{12}^2},
\end{equation}
\begin{equation}\label{theory2:30}
f_{2} \sim \frac{g_{11}\mu_2-g_{12}\mu_1}{g_{11}g_{22}-g_{12}^2}.
\end{equation}
We now define the right hand side of \eqref{theory2:29} to be $n_1$ and the right hand side of \eqref{theory2:30} to be $n_2$, so that $n_j$ is the $j$-th condensate's density far from the vortex. We then rescale the system based on the healing lengths of the condensates, and the `far-off' densities $n_j$. We define the aggregate healing length $\xi \equiv \sqrt{\xi_1 \xi_2}$, where $\xi_j$ is the healing length of the $j$-th condensate. Thus, we identify a new scaled distance parameter $x \equiv \rho / \xi$ and scaled wave function amplitudes $\chi_j = f_j / \sqrt{n_j}$.

We now perform a variational approach on the resulting energy functional,
\begin{dmath}\label{theory2:31}
E = \iint \textrm{d}A \left\{\sum_{j=1,2} \left(\frac{\hbar^2}{2m} \left(\frac{\textrm{d} f_j}{\textrm{d}\rho}\right)^2 + \frac{\hbar^2}{2m}\frac{f_j
^2}{\rho^2}\right) + \frac{g_{11}}{2}f_1^4 + \frac{g_{22}}{2}f_2^4 +g_{12}f_1^2f_2^2\right\}.
\end{dmath}
 Using the rescaled objects defined above, we find that the vortex energy per unit length can be written as
\begin{dmath}\label{theory2:32}
\epsilon_v = \frac{\pi \hbar^2}{m} \int_0^{\frac{D}{\xi}} x dx \left[n_1 \sqrt{n_1 g_{11}n_2 g_{22}}\left(\frac{d\chi_1}{dx}\right)^2 + n_1 \sqrt{n_1 g_{11} n_2 g_{22}}\frac{\chi_1^2}{x^2} + n_2 \sqrt{n_1 g_{11} n_2 g_{22}} \left(\frac{d\chi_2}{dx}\right)^2 + n_2 \sqrt{n_1 g_{11} n_2 g_{22}}\frac{\chi_2^2}{x^2} + \frac{n_1^2 g_{11}}{2}(1-\chi_1^2)^2 + \frac{n_2^2 g_{22}}{2}(1-\chi_2^2)^2 + n_1 n_2 g_{12}(1-\chi_1^2)(1-\chi_2^2)\right],
\end{dmath}
 where $D$ is a large cutoff distance. Applying the trial functions $\chi_j = x/(\alpha_j+x^2)^{1/2}$, which obey $\chi_j\to 1$ as $x \to \infty$, we obtain a minimum at about $\alpha_j = 2$. For this point, we get
\begin{dmath}\label{theory2:33}
\epsilon_v = \frac{\pi \hbar^2}{m}n_1 \left(\frac{1}{2} \sqrt{\frac{n_1 g_{11}}{n_2 g_{22}}}+\log_e{\left(\beta\frac{D}{\xi}\right)}\right)+\frac{\pi \hbar^2}{m}n_2 \left(\frac{1}{2} \sqrt{\frac{n_2 g_{22}}{n_1 g_{11}}}+\log_e{\left(\beta\frac{D}{\xi}\right)}\right) + \frac{\pi \hbar^2}{m} g_{12} \sqrt{\frac{n_1 n_2}{g_{11}g_{22}}},
\end{dmath}
 where $\beta = 0.908$ is a numerical constant. 
 
 As a quick check, let us consider the limit $g_{12}=0$, $n_1=n_2=n$, and $g_{11}=g_{22}$, i.e., a system of two identical non-interacting free condensates. In this case, $\epsilon_v = \frac{2 \pi \hbar^2n}{m} \log_e \left(1.497 \frac{D}{\xi}\right)$, twice the variational result  in Ref. \cite{pethick2002bose}, as expected. This limit also gives us a way to sharpen the numerical constant $\beta$ in Eq. \eqref{theory2:33}. Ref. \cite{pethick2002bose} determines the GPEs for a single free condensate with a vortex, similar to Eqs. \eqref{theory2:27} and \eqref{theory2:28}, computationally to obtain the vortex energy per unit length of $\pi n \frac{\hbar^2}{m} \log_{e} \left(1.464 \frac{D}{\xi}\right)$. We can now propose a more precise value for $\beta$ in Eq. \eqref{theory2:33} to be $0.888$. The expression obtained with this value matches the computational result of Ref. \cite{pethick2002bose} in the required limit. From here on, we therefore use Eq. \eqref{theory2:33} with $\beta = 0.888$ for the vortex energy per unit length.
 
\subsection{\label{sec:theory2:trap}  Harmonic Trap - Static Limit}
We now extend our analysis to the system in the previous section by considering an additional external potential trap of the form $V = \frac{1}{2}m \omega_{\perp} \rho^2$ where $\omega_{\perp}$ is the trap frequency, and we carry out an analysis along the lines of the one conducted in Ref. \cite{pethick2002bose}. Here we assume that the number of atoms in each condensate is large enough so that the Thomas-Fermi approximation is an adequate description of the system. Under these circumstances, in Appendix \ref{subapp:b1} we show that the size of the vortex core, determined by the healing lengths, is much smaller than the radius of each condensate. This is implemented by using the relations
\begin{equation}\label{theory2:35}
\frac{\hbar^2}{2m\xi^2_{0,j}} = n_j (0) g_{jj},
\end{equation}
\begin{equation}\label{theory2:36}
\mu_j = \frac{1}{2}m \omega_{\perp} R_j^2,
\end{equation}
where $\xi_{0,j}$ is the healing length of the $j$-th condensate at the center, $n_j(0)$ is the density of the $j$-th condensate at the center in the absence of rotation, $\mu_j$ is the chemical potential of the $j$-th condensate, and $R_j$ is the radius of the $j$-th condensate. In the absence of ambiguity, we refer to $\xi_{0,j}$ by $\xi_j$ and to $n_j(0)$ by $n_j$. 
 
\begin{figure}
\centering
    \includegraphics[width=0.45\textwidth]{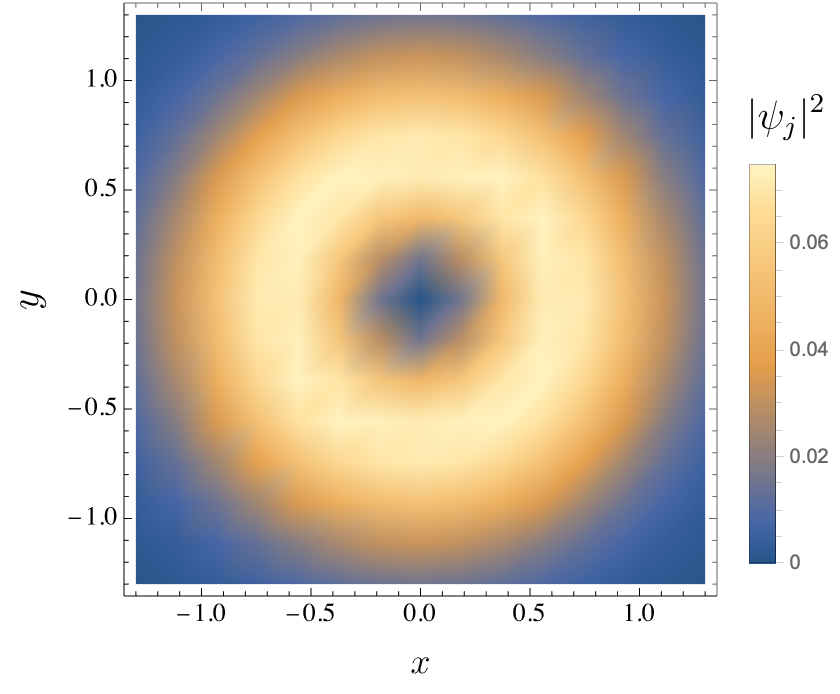}
    \caption{Density plot of the profile of two mixing condensates with a single quantized vortex in a harmonic trap $V=\frac{1}{2}m\omega_{\perp}(\rho^2)$, where $\rho^2 = x^2+y^2$, as considered in Section \ref{sec:theory2:trap}. Here we take the radii of both condensates to be equal. Up to an intermediate distance $\sqrt{\xi_1 \xi_2} \ll \rho_0 \ll R$, where $R$ is the common radius, and $\xi_j$ are the healing lengths of the $j$-th condensate, the radial profile of the system is well approximated by $r/(2+r^2)^{1/2}$, where $r=\rho/\sqrt{\xi_1\xi_2}$. Beyond $\rho_0$, the profile of the system is dictated by hydrodynamic effects from the vortex circulation, i.e., it decays to $0$ as we approach $R$.}
\label{fig:gaussiandip}
\end{figure}

 Up to an intermediate distance $\rho_0$, the vortex energy can be approximated quite well by the results in Appendix \ref{sec:theory2:uni} (Eq. \eqref{theory2:33} particularly), provided $\rho_0$ is much larger than the vortex core size and much smaller than the radius of the condensates, i.e. $\sqrt{\xi_1 \xi_2} \ll \rho_0 \ll \min\{R_1, R_2\}$. At larger distances, while the condensate densities are unaffected by the vortex, the condensates move at speeds determined by the quantized circulation about the vortex. This extra energy can be approximated using a hydrodynamic kinetic term, 
\begin{dmath}\label{theory2:39}
\epsilon_v = \epsilon_u (\rho_0) + \frac{1}{2}m \left(\int_{\rho_0}^{R_1} n_1(r) v^2(r) 2\pi r \, \textrm{d}r + \int_{\rho_0}^{R_2} n_2(r) v^2(r) 2\pi r \, \textrm{d}r \right),
\end{dmath}
where $\epsilon_u(\rho_0)$ is the vortex energy in the uniform limit (as in Eq. \eqref{theory2:33}) evaluated at cutoff distance $\rho_0$, $v(r) = \frac{\hbar}{mr}$ is the hydrodynamic speed, and $n_j (r) = n_j \cdot \left(1 - \frac{(g_{ii} - g_{12})r^2}{g_{ii}R_j^2 - g_{12}R_i^2}\right)$ (again, we use the convention $i \neq j$, and $i, j \in \{1, 2\}$). The expected density profile of the system (in the case that $R_1 = R_2=: R$) is depicted in Fig. \ref{fig:gaussiandip}. In particular, closer to the vortex core, the expected profile looks similar to the uniform case discussed in Section \ref{sec:theory2:uni}, and farther out towards $R$ it exhibits a decay due to hydrodynamic effects.

\begin{figure}
    \centering
    \includegraphics[width=0.5\textwidth]{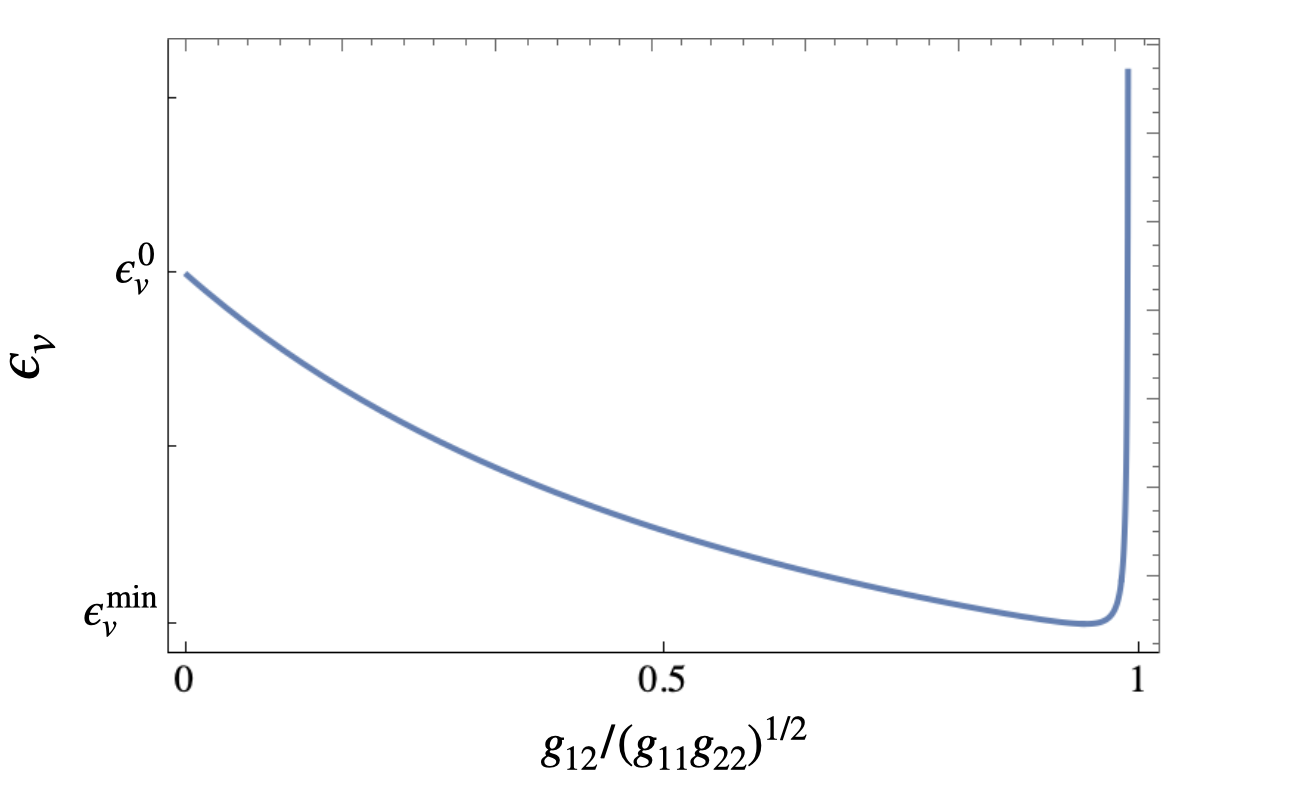}
    \caption{Vortex energy (per unit length) $\epsilon_v$ for a system of two mixing condensates with a singly quantized on-center vortex in a harmonic trap as a function of the scaled interaction parameter $g_{12}/\sqrt{g_{11}g_{22}}$. The value $0$ on the horizontal axis corresponds to the absence of inter-condensate interaction, i.e., the single condensate limit, whereas the value $1$ corresponds to the transition point to the immiscible regime. On the vertical axis, $\epsilon_v^{\mathrm{min}}$ corresponds to the minimal vortex energy of the system, attained just before the transition point. $\epsilon_v^0$ corresponds to the vortex energy in the absence of inter-condensate interaction.}
    \label{fig:energy_g12}
\end{figure}

Evaluating these integrals and adding them to the uniform limit result in the simplifying equal radius case, $R_1 = R_2 =: R$, we find
\begin{equation}\label{theory2:41}
\begin{split}
\epsilon_v &= \frac{\pi \hbar^2}{m}n_1 \left(\frac{1}{2}\sqrt{\frac{n_1 g_{11}}{n_2 g_{22}}} + \log \left(0.539 \frac{R}{\xi}\right)\right)\\ &+ \frac{\pi \hbar^2}{m}n_2 \left(\frac{1}{2}\sqrt{\frac{n_2 g_{22}}{n_1 g_{11}}} + \log \left(0.539 \frac{R}{\xi}\right)\right)\\ &+ \frac{\pi \hbar^2}{m}g_{12} \sqrt{\frac{n_1 n_2}{g_{11}g_{22}}}.
\end{split}
\end{equation}
The integrals in Eq.\eqref{theory2:39} can also be evaluated fairly easily when $R_1 \neq R_2$, and the results are presented in Appendix \ref{subapp:b2} - these details have been omitted from this section for ease of presentation. A plot of $\epsilon_v$ over varying $g_{12}/\sqrt{g_{11}g_{22}}$ values is shown in Fig. \ref{fig:energy_g12}. The most notable feature of this plot is the divergence observed at the transition point $g_{12} = \sqrt{g_{11}g_{22}}$, and the behavior of the system right before. In particular, the minimal vortex energy (per unit length) is attained right before the transition. This feature is discussed in Section \ref{sec:discussion2}.

We now make a brief note regarding off-axis vortices, located at a radius $b$ from the center in the equal radius setup. In this case, the condensate densities scale like $\left(1- \frac{b^2}{R^2}\right)$, and we find that $\epsilon_v (b) = \left(1 - \frac{b^2}{R^2}\right) \epsilon_v$, where $\epsilon_v$ is the result in Eq. \eqref{theory2:41}. We also compute the total angular momentum of the system, which is given by
\begin{equation}\label{theory2:42}
\mathcal{L} = \hbar \left(\int_0^{R_1} n_1(r) \, \mathrm{d}r + \int_0^{R_2} n_2 (r) \, \mathrm{d}r\right).
\end{equation}
Again, in the equal radius setup (see Appendix \ref{subapp:b2} for the general unequal radius result), this quantity is evaluated to be
\begin{equation}\label{theory2:44}
\mathcal{L} = \frac{\pi \hbar}{2} (n_1 + n_2) R^2.
\end{equation}
For an off-axis vortex at a radius $b$ in the equal radius system, we similarly evaluate the angular momentum, 
\begin{equation}\label{theory2:45}
\mathcal{L}(b) = \frac{\pi \hbar}{2} (n_1 + n_2)\left(R^2 - 2b^2 + \frac{b^4}{R^2}\right).
\end{equation}
For $b \ll R$, this simplifies to
\begin{equation}\label{theory2:46}
\mathcal{L}(b) = \frac{\pi \hbar R^2}{2} (n_1 + n_2)\left(1 - \frac{2b^2}{R^2}\right).
\end{equation}
These quantities will be helpful when we compute the critical and metastability onset angular velocities in Section \ref{sec:theory2:rottrap} (along the lines of Ref. \cite{pethick2002bose}). 

\subsection{\label{sec:theory2:rottrap} Rotating Harmonic Trap}
Next consider the case where we rotate the trap in Section \ref{sec:theory2:trap} with an angular velocity $\Omega$. Suppose the state with angular momentum $L$ has energy $E_L$. This state will be energetically favorable compared to the ground state $E_0$ if $\Omega$ exceeds a critical value $\Omega_c$, 
\begin{equation}\label{theory2:47}
\Omega_c = \frac{E_L - E_0}{L}
\end{equation}
For a single vortex in the Thomas-Fermi regime, we compute $E_L - E_0$ and $L$ in Appendix \ref{app:b} (Eqs. \eqref{theory2:40} and \eqref{theory2:43}). Using the equal radius assumption, we find that the \emph{critical angular velocity} $\Omega_c$ is given by
\begin{dmath}\label{theory2:48}
\Omega_c = \frac{2\hbar}{m (n_1 + n_2)R^2}\left(\frac{n_1}{2} \sqrt{\frac{n_1 g_{11}}{n_2 g_{22}}}+\frac{n_2}{2}\sqrt{\frac{n_2 g_{22}}{n_1 g_{11}}}+g_{12} \sqrt{\frac{n_1 n_2}{g_{11}g_{22}}} + (n_1+n_2) \log\left(0.539 \frac{R}{\xi}\right)\right).
\end{dmath}

To find the angular velocity $\Omega_m$ at which a vortex achieves \emph{metastability} or \emph{local stability} at the center of the trap, we first compute the energy of an off-axis vortex in the equal radius system using the rotating frame. In the language of \cite{Fetter_2001,Fetter_2008}, we will refer to $\Omega_m$ as the \emph{metastability onset angular velocity}. The energy of an off-axis vortex in the rotating frame is given by $\tilde{E} = E - \mathbf{\Omega} \cdot \mathbf{L}$. 

\begin{figure}[H]
\centering
\begin{subfigure}[b]{0.47\textwidth}
    \centering
    \includegraphics[width=\textwidth]{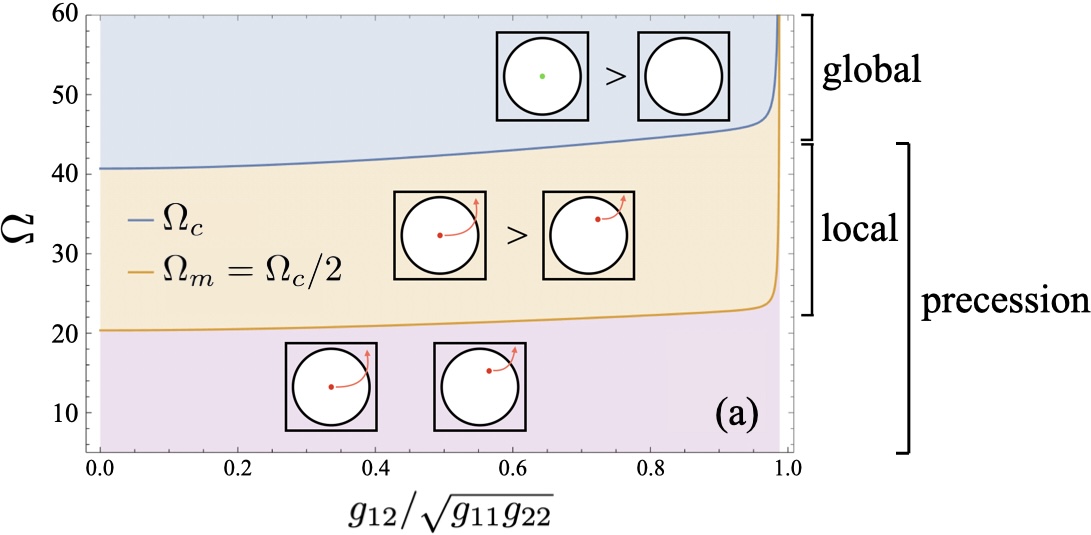}
    \label{fig:omegac_graph}
\end{subfigure}

\begin{subfigure}[b]{0.38\textwidth}
\centering
\includegraphics[width=\textwidth]{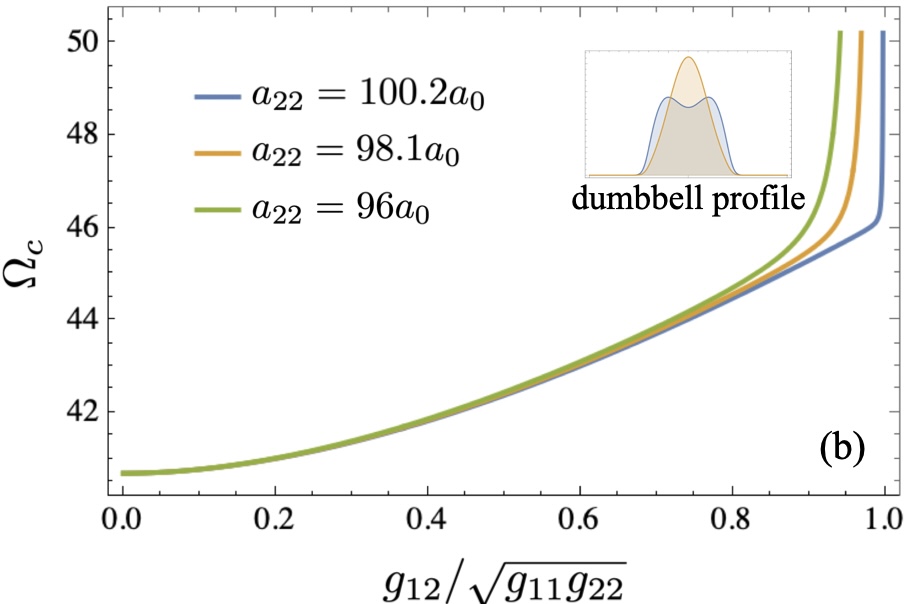}
\label{fig:dumbbellplots}
\end{subfigure}
\caption{(a) Phase diagram of a single vortex in a binary condensate system marked with schematics for the various regimes. The vertical axis represents $\Omega$, the angular velocity of the rotating harmonic trap, and the horizontal axis represents the scaled interaction parameter $g_{12}/\sqrt{g_{11}g_{22}}$. $\Omega_c$ demarcates the zones of \emph{global stability} (marked ``global,"   blue tier in the figure) and \emph{precession}. For $\Omega>\Omega_c$, an on-center vortex is globally stable, i.e., an on-center vortex state is energetically preferred over the no-vortex state. If $\Omega < \Omega_c$, any vortex formed within the system eventually precesses out of the condensates. The region $\Omega_m < \Omega < \Omega_c$ represents the zone of \emph{local stability} (marked ``local,''  orange tier in the figure). Within this region, an on-center vortex state is energetically preferred over an off-center vortex state, but is not preferred over a no-vortex state.
(b) Dependence of the critical angular velocity $\Omega_c$ of the binary system on the scaled interaction parameter $g_{12}/\sqrt{g_{11}g_{22}}$ for three different values of $a_{22}$, the scattering length of the second condensate (or equivalently, the intra-interaction parameter $g_{22}$ of the second condensate), fixing $a_{11}$ at $100.4a_0$. The figure shows a divergence point shifting towards the transition point at $g_{12} = \sqrt{g_{11}g_{22}}$, as the difference between $a_{11}$ and $a_{22}$ (or equivalently  between $g_{11}$ and $g_{22}$) decreases. The inset captioned ``dumbbell profile'' is a qualitative depiction of the radial profile of the system after the divergence point is reached, in a phase we refer to in Section \ref{sec:discussion2} as the \emph{dumbbell phase}. The condensate colored blue in the inset is the one with a larger effective strength parameter $Ng_{ii}$, and it exhibits a characteristic bimodal distribution.}
\label{fig:omegacplots}
\end{figure}

Using results from Section \ref{sec:theory2:trap},
\begin{equation}\label{theory2:49}
\tilde{E}(b) = \epsilon_v \cdot \left(1 - \frac{b^2}{R^2}\right) - \Omega \mathcal{L} \cdot \left(1 - \frac{2b^2}{R^2}\right),
\end{equation} 
\noindent where $b$ is the distance of the vortex from the center (and we assume $b \ll R$), and $\epsilon_v$ and $\mathcal{L}$ are the quantities in Eqs. \eqref{theory2:41} and \eqref{theory2:44}. We observe that $\tilde{E}(b)$ has a local minimum at $b=0$ iff $\Omega \geq \Omega_c/2$, i.e. for $\Omega \geq \Omega_c/2$ we have $\tilde{E}'(b)=0$ and $\tilde{E}''(b) \geq 0$. So for angular velocities above $\Omega_c/2$, a vortex at the center is locally stable. We conclude that the metastability onset angular velocity $\Omega_m$ is indeed $\Omega_c/2$.

The results of this section are visually compiled in Fig. \ref{fig:omegacplots}. In Fig. \ref{fig:omegacplots} (a) we plot the cutoff velocities $\Omega_c$ and $\Omega_m$,  providing a phase diagram for the stability of a single vortex state in the system. The contents of this phase diagram can be summarized as follows. For $\Omega<\Omega_c$, any vortex formed will precess out of the condensates eventually. Beyond $\Omega_c$, however, a state with an on-center vortex is energetically preferred to a no-vortex state--that is to say, on-center vortices are globally stable for angular velocities above  $\Omega_c$. In the intermediate regime between $\Omega_m$ and $\Omega_c$, an on-center vortex state is energetically preferred to an off-center vortex state, i.e., on-center vortices are locally stable for $\Omega_m < \Omega < \Omega_c$. 

In Fig. \ref{fig:omegacplots} (b), we demonstrate the effect of varying the separation between the intra-atomic scattering lengths on the behavior of $\Omega_c$ with respect to the scaled inter-interaction $g_{12}/\sqrt{g_{11}g_{22}}$. There are two most notable features in this plot. First, in each case there is a strong divergence of $\Omega_c$ that occurs before the transition point at $g_{12}=\sqrt{g_{11}g_{22}}$. Second, this divergence point of $\Omega_c$ appears to move closer to the transition point as the gap between the intra-atomic scattering lengths narrows. The behavior observed in Fig. \ref{fig:omegacplots} (b) is explained in Section \ref{sec:discussion2}.

\begin{table}
    \centering
    \begin{tabular}{|p{2cm}|p{3cm}|p{2cm}|}
\hline
\multicolumn{3}{|c|}{Shooting Method Pseudoparameters}\\
\hline
Parameter & Definition & Test Values\\
\hline
     $a_{1}$ & \footnotesize $\sqrt{(n_1g_{11})/(n_2g_{22}})$ & 1\\
     $a_{2}$ & \footnotesize $(g_{12}/g_{22}) \cdot (1/a_{1})$ & 0.1\\
     $a_3$& \footnotesize $a_1 + (g_{12})/(g_{11}a_1)$ & 1\\
     $b_1$& \footnotesize $1/a_1$ & 1\\
     $b_2$ & \footnotesize $(g_{12}a_1)/g_{11}$ & 0.1\\
     $b_3$ & \footnotesize $1/a_1+(g_{12}a_1)/g_{22}$ & 1\\
\hline
\end{tabular}
    \caption{Table of pseudoparameters used for the shooting method calculations in Section \ref{sec:comp2}. The first column provides the (pseudo)parameter, the second column  its definition in terms of quantities introduced in Sections \ref{sec:nonrot} and \ref{sec:theory2:uni}, and the final column provides test values used to plot the results shown in Fig. \ref{fig:shoot}. Particularly, $n_j$  is the $j$-th condensate's density far from the vortex, $g_{jj}$ is the intra-condensate interaction parameter of the $j$-th condensate, and $g_{ij}$ (for $i \neq j$) is the inter-condensate interaction parameter.}
    \label{table:tab1}
\end{table}

\subsection{\label{sec:comp2} Computational Analysis}
\noindent In this Section we outline the use of the shooting method (see \cite{Pang_2006} for an introduction) to verify the computational accuracy of the ansatz used in Section \ref{sec:theory2:uni}. We first introduce a set of pseudoparameters listed in Table \ref{table:tab1}.

We would like to solve the following system of ODEs, with boundary conditions $\chi_i(0)=0$ and $\chi_i(\infty)=1$,
\begin{equation}\label{comp:50-51}
    \begin{split}
        -\frac{1}{x} \frac{d}{dx} \left(x \frac{d \chi_1}{dx}\right) + \frac{\chi_1}{x^2} + a_1\chi_1^3
        + a_2 \chi_1 \chi_2^2 &= a_3 \chi_1,\\
        -\frac{1}{x} \frac{d}{dx} \left(x \frac{d \chi_2}{dx}\right) + \frac{\chi_2}{x^2} + b_1\chi_2^3 + b_2 \chi_2 \chi_1^2 &= b_3 \chi_2 .
    \end{split}
\end{equation}

\begin{figure}
    \centering
    \includegraphics[width=0.45\textwidth]{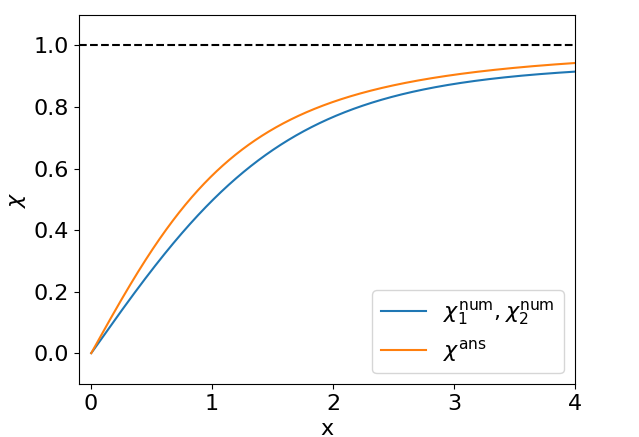}
    \caption{Numerically obtained (scaled) radial profiles $\chi^{\mathrm{num}}_{i}$ of a system of binary condensates in the uniform limit with a singly quantized on-center vortex, compared with the variational ansatz $\chi^{\mathrm{ans}}(x)=x/(2+x^2)^{1/2}$ proposed in Section \ref{sec:theory2:uni}. The quantity $x$  on the horizontal axis of the plot is the radial distance $\rho$ scaled by the \emph{aggregate healing length} $\xi=\sqrt{\xi_1\xi_2}$. The profiles $\chi^{\mathrm{num}}_{i}$ were obtained applying the shooting method  detailed in Section \ref{sec:comp2} over $N=10000$ steps on the interval $u=10^{-3}$ to $u=0.95$, using the values for the pseudoparameters listed in Table \ref{table:tab2}. Here $u$ is the scaled distance $u=x/(1+x)$ used to enforce the boundary conditions at $x=0$ and $x=\infty$. This figure illustrates the proximity between the numerical solutions and the variational ansatz.}
    \label{fig:shoot}
\end{figure}

When changing variables, $u= x/(1+x)$, the boundary conditions transform to $\chi_i(u=0) = 0$ and $\chi_i(u=1)=1$. Furthermore, we define $y_i = \frac{d\chi_i}{dx}$. The shooting method can then be applied to the system of ODEs obtained by rewriting Eq. \eqref{comp:50-51} using these redefinitions (see also Appendix \ref{subapp:b3}). Fig. \ref{fig:shoot} shows the numerically obtained results (taking the generic pseudoparameter values in Table \ref{table:tab1} as input), in comparison with the variational solution from Section \ref{sec:theory2:uni}. It serves as a concrete illustration that the variational solution obtained is reasonably accurate.

\subsection{\label{sec:discussion2} Discussion}
In this section, we analyze the results of Section \ref{sec:theory2:rottrap}, and specifically the divergences observed in Fig. \ref{fig:omegacplots} (b). The subfigure shows plots of $\Omega_c$ against the inter-condensate interaction for narrow and wide separations between $g_{11}$ and $g_{22}$. As mentioned earlier, the first remarkable aspect of the plot are the stronger-than-polynomial divergences, all of which occur before the transition point at $g_{12} = \sqrt{g_{11}g_{22}}$. What we see is that as the difference between the intra-condensate interactions (quantified as $g_{11}, g_{22}$) widens, the divergence point appears earlier as we increase the inter-condensate interaction; and as the difference between the intra-interactions narrows, the divergence point appears later. In the limit  $g_{11}=g_{22}$, this divergence appears exactly at the transition point.

This divergent behavior can be explained as follows. First consider the density profile of the system in the miscible phase in the absence of rotation (see left side of Fig. \ref{fig:delta v w} for the characteristic 2D profile in this regime). For simplicity, we may further restrict our renditions going forth in this section to the 1D profiles described in Ref. \cite{wirthwein2022collision} These profiles can be visualized as two concentric inverted parabolas, with slightly different peak heights. As the inter-interaction $g_{12}$ is increased toward the transition point, one sees that the weaker condensate (that is to say the condensate with smaller $Ng_{ii}$) holds the center, whereas the stronger condensate (with larger $Ng_{ii}$) begins to split into a bimodal density distribution about the central condensate (shown in the inset in Fig. \ref{fig:omegacplots} (b)). While this is still in the mixing phase (since the centers of both condensates are aligned), the formation of a bimodal state indicates a new subregime that has consistently been observed in the ground states of binary $^{87}\mathrm{Rb}$-$^{87}\mathrm{Rb}$ condensate systems. We term this sub-regime the \emph{dumbbell phase}. 

The physical analogy is as follows: imagine spinning a bowling ball around its center. This resembles the effort an on-center vortex requires to remain globally stable at lower values of $g_{12}$. Now, as the bowling ball gradually splits into a dumbbell shape, with its axle extending outward, continuing to spin it about the axle's center demands significantly more effort—effort that increases as the axle extends further. Similarly, in the dumbbell phase, an on-center vortex requires a much larger value of $\Omega_c$ to maintain global stability. In precise terms, the variance of the condensate distribution—analogous to the moment of inertia of the system about an axis through the trap center—diverges as the system transitions to the dumbbell phase. This divergence is reflected in the behavior of (b) in Fig. \ref{fig:omegacplots}.

Before accepting this analogy, one might ask how the entry point to the dumbbell phase varies with wider and narrower gaps between the intra-interactions as in Fig. \ref{fig:omegacplots}. We see that for 1D systems, as the gap between the intra-interactions $g_{11}$ and $g_{22}$ is narrowed down to near 0, there is virtually no intervening phase!  Similarly, as this gap is widened, the dumbbell phase appears earlier as we increase the $g_{12}$ interaction. The remarkable result from this analogy is that we have been able to observe a novel phase in the $^{87}\mathrm{Rb}$-$^{87}\mathrm{Rb}$ condensate system (even in non-rotating contexts), purely informed by vortex formation and stability in the rotating trapped system. 

\section{\label{sec:conclusion} Conclusions}
\noindent In this work, we have discussed the formation and stability of singly quantized vortices in a system of binary BECs. We began in Section \ref{sec:nonrot} with two non-rotating condensates to illustrate the vortex-free theory. Here we employed a variational analysis of this system to note a transition between a \emph{miscible} and an \emph{immiscible} phase as one scales the inter-condensate interaction parameter $g_{12}$. In Section \ref{sec:rot} we performed a step-by-step analysis of a binary system of interacting rotating BECs. We began in Section \ref{sec:theory2:uni} by outlining the theory of a binary BEC system in the uniform limit with a singly quantized on-center vortex. In particular, we derived a form for the asymptotic densities of the condensates and an approximate form for the vortex energy of the system. In Section \ref{sec:theory2:trap}, we placed the system in a harmonic trap and derived forms for the vortex energy of the system, as well as the total angular momentum of the system. Here, we also provided arguments for the vortex energy and total system angular momentum associated to an off-center vortex at a distance $b$ away from the center. In Section \ref{sec:theory2:rottrap} we rotated the trap with an angular velocity $\Omega$ and analyzed vortex stability regimes (and the corresponding cutoff angular velocities) that are observed for single condensate systems in classical literature (see \cite{bao_2006, Wieman_2000, Coddington_2004, Fetter_2008, Fetter_2001, Kavoulakis_2000, Madison_2000} for examples). We checked our work in Section \ref{sec:theory2:uni} computationally using the shooting method in Section \ref{sec:comp2}, and the results were remarkably close to the ansatz used in our analysis. Finally, in Section \ref{sec:discussion2} we discussed the presence of a novel \emph{dumbbell phase} for binary BECs - with or without vortices -  informed by our analysis of vortex formation and stability in the system. It is remarkable that this phenomenon has not yet been noted in earlier work until examining the vortex stability in the system whereby it naturally manifests itself.

It remains to concretely understand what quantitative parameters govern the dumbbell phase. As mentioned in Section \ref{sec:discussion2}, returning to mechanical intuition, one way to do this is to numerically obtain the moment of inertia of the system (which in our context is simply the variance of the condensate distributions) and check for divergences as we tune the interaction parameter $g_{12}$. At the moment, we do not have a tunable (numerical) form for the condensate wave functions in Section \ref{sec:nonrot}, and hence obtaining the moment of inertia of the system is out of our reach as of yet. It is also certainly possible that there is a simple quantitative indicator of bimodality in a distribution that makes the presence of a dumbbell phase apparent, providing an order parameter for this phase.

We conclude with remarks on some of the work done in \cite{Kasamatsu2003, Doran2024, daSilva_2023} and how it relates to the framework employed here. Refs. \cite{Kasamatsu2003} and \cite{Doran2024} are perhaps closest in perspective to ours, filling in aspects of the theory of vortex lattices and immiscible systems that we have not discussed here. In \cite{Kasamatsu2003}, the authors establish a phase diagram for vortex lattices in binary Bose-Einstein condensates over various values of the inter-component coupling. They show that in the immiscible regime, which we have explicitly avoided in later sections of this article, vortex lattices evolve into double-core vortex lattices and then into interwoven vortex sheets at high enough $\Omega$. \cite{Doran2024} explores binary BEC systems in the immiscible regime with a majority component that contains a singly quantized vortex and a minority component that fills the vortex core. The authors are able to numerically and variationally calculate the steady state of the in-filled vortex and establish its stability under perturbation of the in-filling component away from the core. \cite{daSilva_2023} discusses yet another important aspect of the theory, the dynamics of vortex formation in binary systems. While we have restricted ourselves to binary BEC systems with identical components here, \cite{daSilva_2023} is devoted to mass-imbalanced binary BECs. It studies the time evolution of the vortex states in these systems introduced by a stirring trap mechanism, emphasizing the effect of mass imbalances in the dynamics.
\acknowledgements
The authors are grateful to Abhinav Prem for useful communications, and acknowledge the Center for Advanced Research Computing (CARC) at the University of Southern California for providing computing resources that have contributed to the research results reported within this publication. URL: \href{https://carc.usc.edu}{https://carc.usc.edu}.

\appendix
\section{Variational Analysis in Section  \ref{sec:nonrot}}\label{app:a}
In this appendix, we present key details of the variational analysis within the setup considered in Section \ref{sec:nonrot}, leading to the expressions for $\sigma_x$, $\sigma_y$, and $\delta$ in Eqs. \eqref{theory1:8}, \eqref{theory1:9}, and \eqref{theory1:10}. 

Beginning with the evaluated energy functional in Eq. \eqref{theory1:6}, we differentiate with respect to the variational parameters to obtain the ground state conditions for the system,
\begin{dmath}\label{a:1}
    -\frac{\hbar^2}{2m\sigma_{1x}^4} - \frac{Ng_{11}}{4\pi \sigma_{1y}\sigma_{1x}^3} + \frac{1}{2} m \omega_x^2  = \frac{Ng_{12}(\sigma_{1x}^2 + \sigma_{2x}^2-2\delta_x^2)e^{-\frac{\delta_x^2}{\sigma_{1x}^2+\sigma_{2x}^2}-\frac{\delta_y^2}{\sigma_{1y}^2+\sigma_{2y}^2}}}{\pi (\sigma_{1x}^2+\sigma_{2x}^2)^2\sqrt{(\sigma_{1x}^2+\sigma_{2x}^2)(\sigma_{1y}^2+\sigma_{2y}^2)}},
\end{dmath}
\begin{dmath}\label{a:2}
     -\frac{\hbar^2}{2m\sigma_{2x}^4} - \frac{Ng_{22}}{4\pi \sigma_{2y}\sigma_{2x}^3} + \frac{1}{2} m \omega_x^2 = \frac{Ng_{12}(\sigma_{1x}^2 + \sigma_{2x}^2-2\delta_x^2)e^{-\frac{\delta_x^2}{\sigma_{1x}^2+\sigma_{2x}^2}-\frac{\delta_y^2}{\sigma_{1y}^2+\sigma_{2y}^2}}}{\pi (\sigma_{1x}^2+\sigma_{2x}^2)^2\sqrt{(\sigma_{1x}^2+\sigma_{2x}^2)(\sigma_{1y}^2+\sigma_{2y}^2)}},
\end{dmath}
\begin{dmath}\label{a:3}
    -\frac{\hbar^2}{2m\sigma_{1y}^4} - \frac{Ng_{11}}{4\pi \sigma_{1y}^3\sigma_{1x}} + \frac{1}{2} m \omega_y^2 = \frac{Ng_{12}(\sigma_{1y}^2 + \sigma_{2y}^2-2\delta_y^2)e^{-\frac{\delta_x^2}{\sigma_{1x}^2+\sigma_{2x}^2}-\frac{\delta_y^2}{\sigma_{1y}^2+\sigma_{2y}^2}}}{\pi (\sigma_{1y}^2+\sigma_{2y}^2)^2\sqrt{(\sigma_{1x}^2+\sigma_{2x}^2)(\sigma_{1y}^2+\sigma_{2y}^2)}},
\end{dmath}
\begin{dmath}\label{a:4}
    -\frac{\hbar^2}{2m\sigma_{2y}^4}-\frac{Ng_{22}}{4\pi \sigma_{2y}^3\sigma_{2x}} + \frac{1}{2} m \omega_y^2  = \frac{Ng_{12}(\sigma_{1y}^2 + \sigma_{2y}^2-2\delta_y^2)e^{-\frac{\delta_x^2}{\sigma_{1x}^2+\sigma_{2x}^2}-\frac{\delta_y^2}{\sigma_{1y}^2+\sigma_{2y}^2}}}{\pi (\sigma_{1y}^2+\sigma_{2y}^2)^2\sqrt{(\sigma_{1x}^2+\sigma_{2x}^2)(\sigma_{1y}^2+\sigma_{2y}^2)}},
\end{dmath}
\begin{equation}\label{a:5}
    \small \delta_x \left(\frac{4Ng_{12} e^{-\frac{\delta_x^2}{\sigma_{1x}^2+\sigma_{2x}^2}-\frac{\delta_y^2}{\sigma_{1y}^2+\sigma_{2y}^2}}}{\pi (\sigma_{1x}^2+\sigma_{2x}^2)\sqrt{(\sigma_{1x}^2+\sigma_{2x}^2)(\sigma_{1y}^2+\sigma_{2y}^2)}}-m \omega_x^2\right) = 0,
\end{equation}
\begin{equation}\label{a:6}
    \small \delta_y \left(\frac{4Ng_{12} e^{-\frac{\delta_x^2}{\sigma_{1x}^2+\sigma_{2x}^2}-\frac{\delta_y^2}{\sigma_{1y}^2+\sigma_{2y}^2}}}{\pi (\sigma_{1y}^2+\sigma_{2y}^2)\sqrt{(\sigma_{1x}^2+\sigma_{2x}^2)(\sigma_{1y}^2+\sigma_{2y}^2)}} - m \omega_y^2\right) = 0.
\end{equation}

Applying the balance relation Eq. \eqref{theory1:7} in Section \ref{sec:nonrot} to the ground state conditions above, we obtain the  simplifications
\begin{equation}\label{a:7}
    \sigma_x^4 = \frac{N g_{12} \omega_y e^{-\frac{\delta_x^2}{2\sigma_x^2}-\frac{\delta_y^2}{2\sigma_y^2}}}{\pi m \omega_x^3},
\end{equation}
\begin{equation}\label{a:8}
    \sigma_y^4 = \frac{N g_{12} \omega_x e^{-\frac{\delta_x^2}{2\sigma_x^2}-\frac{\delta_y^2}{2\sigma_y^2}}}{\pi m \omega_y^3},
\end{equation}
\begin{equation}\label{a:9}
   \small -\frac{\hbar^2}{2m\sigma_x^4}-\frac{N\omega_y g}{4\pi \sigma_x^4 \omega_x}+\frac{1}{2}m \omega_x^2 = \frac{Ng_{12}\omega_y(\sigma_x^2 - \delta_x^2)e^{-\frac{\delta_x^2}{2\sigma_x^2}-\frac{\delta_y^2}{2\sigma_y^2}}}{4\pi \sigma_x^6 \omega_x},
\end{equation}
\begin{equation}\label{a:10}
    \small -\frac{\hbar^2}{2m\sigma_y^4}-\frac{N\omega_x g}{4\pi \sigma_y^4 \omega_y}+\frac{1}{2}m \omega_x^2 = \frac{Ng_{12}\omega_x(\sigma_y^2 - \delta_y^2)e^{-\frac{\delta_x^2}{2\sigma_x^2}-\frac{\delta_y^2}{2\sigma_y^2}}}{4\pi \sigma_y^6 \omega_y}.
\end{equation}

To simplify the situation further, as mentioned in Section \ref{sec:nonrot}, we drop the kinetic terms from Eqs. \eqref{a:9} and \eqref{a:10} (this is consistent with the large $N$ limit). We also restrict ourselves to separation only along the x-direction, so that $\delta_y=0$ and $\delta := \delta_x \geq 0$. This resolves the ground state conditions further to
\begin{equation}\label{a:11}
    \sigma_x^4 = \frac{N g_{12} \omega_y e^{-\frac{\delta^2}{2\sigma_x^2}}}{\pi m \omega_x^2},
\end{equation}
\begin{equation}\label{a:12}
    \sigma_y^4 = \frac{N g_{12} \omega_x e^{-\frac{\delta^2}{2\sigma_x^2}}}{\pi m \omega_y^2},
\end{equation}
\begin{equation}\label{a:13}
    -\frac{N\omega_y g}{4\pi \sigma_x^4 \omega_x}+\frac{1}{2}m \omega_x^2 = \frac{Ng_{12}\omega_y(\sigma_x^2 - \delta^2)e^{-\frac{\delta^2}{2\sigma_x^2}}}{4\pi \sigma_x^6 \omega_x},
\end{equation}
\begin{equation}\label{a:14}
    -\frac{N\omega_x g}{4\pi \sigma_y^4 \omega_y}+\frac{1}{2}m \omega_y^2 = \frac{Ng_{12}\omega_x(\sigma_y^2)e^{-\frac{\delta^2}{2\sigma_x^2}}}{4\pi \sigma_y^6 \omega_y}.
\end{equation}
For  notational convenience, we also set $\mathcal{G}:=e^{-\frac{\delta^2}{2\sigma_x^2}}$. We now focus on Eqs. \eqref{a:11} and \eqref{a:13}. The condition Eq. \eqref{a:13} simplifies to
\begin{equation}\label{a:15}
\sigma_x^4 = \frac{N \omega_y}{2 m \pi \omega_x^3} \left(g+\frac{g_{12}(\sigma_x^2 - \delta^2)\mathcal{G}}{\sigma_x^2}\right).
\end{equation}
Using Eq. \eqref{a:11}, we obtain the expression
\begin{equation}\label{a:16}
g/g_{12} = \left(1+\frac{\delta^2}{\sigma_x^2}\right) \mathcal{G} \equiv \left(1+\frac{\delta^2}{\sigma_x^2}\right) e^{-\frac{\delta^2}{2\sigma_x^2}}.
\end{equation}

Finally, Eq. \eqref{a:16} may be Taylor expanded for small $\delta$, and using Eqs. \eqref{a:11} and \eqref{a:12} the low-order approximations yield approximations for $\delta$, $\sigma_x$, and $\sigma_y$. 
\begin{equation}\label{a:17}
\sigma_x^4 = \frac{N\omega_y}{m\pi \omega_x^3} \frac{g+g_{12}}{2},
\end{equation}
\begin{equation}\label{a:18}
\sigma_y^4 = \frac{N\omega_x}{m\pi \omega_y^3} \frac{g+g_{12}}{2},
\end{equation}
\begin{equation}\label{a:19}
\delta = \left(\frac{2N\omega_y}{m\pi \omega_x^3}(g+g_{12})\right)^{1/4} \sqrt{\log_{e} \left(\frac{g_{12}}{g}\right)}.
\end{equation}

\section{Details of Rotating Condensates in Section \ref{sec:rot}}\label{app:b}
This appendix is a supplement to Section \ref{sec:rot}, presenting key details omitted from its subsections for presentation purposes.
\subsection{Vortex Core Sizes}\label{subapp:b1}
In this subsection, we show that the core size of the vortex considered in Section \ref{sec:theory2:trap} (determined by the healing lengths of the condensates) is much smaller than the radii of the condensates (restricting ourselves to the Thomas-Fermi regime of the system). Recall first the relation
\begin{equation}\label{theory2:35}
\frac{\hbar^2}{2m\xi^2_{0,j}} = n_j (0) g_{jj},
\end{equation}
where $\xi_{0,j}$ is the healing length of the $j$-th condensate ($j=1,2$) at the center, and $n_j (0)$ is the density of the $j$-th condensate at the center in the absence of rotation. As in Section \ref{sec:theory2:trap}, we call these quantities $\xi_j$ and $n_j$, provided there is no concern of ambiguity. The chemical potential of the $j$-th condensate ($\mu_j$) is related to the trap frequency $\omega_{\perp}$, and the radius of the $j$-th condensate ($R_j$) by the relation
\begin{equation}\label{theory2:36}
\mu_j = \frac{1}{2}m \omega_{\perp} R_j^2.
\end{equation}
\noindent From Section \ref{sec:theory2:uni}, we also know
\begin{equation}\label{theory2:37}
n_j g_{jj} = \frac{\mu_j - \frac{g_{12}}{g_{ii}}\mu_i}{1-\frac{g_{12}^2}{g_{11}g_{22}}},
\end{equation}
where $i \neq j$ and $i, j \in \{1,2\}$. This leads to the relation
\begin{equation}\label{theory2:38}
\frac{\xi_j}{R_j}=\frac{\hbar \omega_{\perp}}{2 \mu_j} \sqrt{\frac{1-\frac{g_{12}^2}{g_{11}g_{22}}}{1-\frac{g_{12}}{g_{ii}}\frac{R_i^2}{R_j^2}}}.
\end{equation}
In the Thomas-Fermi regime, the right hand side of Eq. \eqref{theory2:38} is  small (since the chemical potential $\mu_j$ is much larger than the trap energy $\hbar \omega_{\perp}$), so we conclude that $\xi_j \ll R_j$ in this setup, as desired.
\subsection{Unequal Radius Results}\label{subapp:b2}
In this subsection, we present results from the analysis in Section \ref{sec:theory2:trap} and Section \ref{sec:theory2:rottrap} for unequal radius condensates, that is, condensates such that $R_1$ and $R_2$ are allowed to be distinct.

By evaluating the integrals in Eq. \eqref{theory2:39} for a generic pair $(R_1, R_2)$, we obtain the vortex energy per unit length,
\begin{dmath}\label{theory2:40}
\epsilon_v = \frac{\pi \hbar^2}{m}n_1 \left(\frac{1}{2}\sqrt{\frac{n_1 g_{11}}{n_2 g_{22}}} + \log \left(0.888 \frac{R_1}{\xi}\right) - \frac{(g_{12}-g_{22})R_1^2}{2(g_{12}R_2^2-g_{22}R_1^2)}\right)\\ + \frac{\pi \hbar^2}{m}n_2 \left(\frac{1}{2}\sqrt{\frac{n_2 g_{22}}{n_1 g_{11}}} + \log \left(0.888 \frac{R_2}{\xi}\right) - \frac{(g_{12}-g_{11})R_2^2}{2(g_{12}R_1^2-g_{11}R_2^2)}\right)\\ + \frac{\pi \hbar^2}{m}g_{12} \sqrt{\frac{n_1 n_2}{g_{11}g_{22}}}.
\end{dmath}

Similarly, by evaluating the integrals in Eq. \eqref{theory2:42} for a generic pair $(R_1, R_2)$, obtain the angular momentum of the system to be
\begin{dmath}\label{theory2:43}
\mathcal{L} = \frac{\pi \hbar}{2} \left(2n_1R_1^2 + n_1\frac{g_{12}-g_{22}}{g_{22}R_1^2-g_{12}R_2^2}R_1^4 + 2n_2R_2^2 + n_2\frac{g_{12}-g_{11}}{g_{11}R_2^2-g_{12}R_1^2}R_2^4\right).
\end{dmath}

As in Section \ref{sec:theory2:rottrap}, this gives us enough information to compute the critical angular velocity $\Omega_c$ when $R_1 \neq R_2$. The result of this, of course, is just the quotient of Eq. \eqref{theory2:40} by Eq. \eqref{theory2:43}. 

\subsection{Shooting Method}\label{subapp:b3}
In this subsection, we present the system of ODEs obtained by rewriting the GPEs in Eq. \eqref{comp:50-51} after the change of variables enforcing the boundary conditions (see Section \ref{sec:comp2} for these redefinitions). The system of ODEs thus obtained is 
\begin{equation}\label{comp:52}
    \begin{split}
        \frac{d\chi_i}{du} &= \frac{y_i}{(1-u)^2}\\
        \frac{dy_1}{du} &= \frac{a_1 \chi_1^3}{(1-u)^2} + \frac{a_2}{(1-u)^2}\chi_1\chi_2^2 - \frac{y_1}{u(1-u)}\\ &+ \frac{\chi_1}{u^2} - \frac{a_3 \chi_1}{(1-u)^2}\\
        \frac{dy_2}{du} &= \frac{b_1 \chi_2^3}{(1-u)^2} + \frac{b_2}{(1-u)^2}\chi_2\chi_1^2 - \frac{y_2}{u(1-u)}\\ &+ \frac{\chi_2}{u^2} - \frac{b_3 \chi_2}{(1-u)^2}
    \end{split}
\end{equation}
where $i \in \{1,2\}$. The results from the analysis presented in Section \ref{sec:comp2} are now obtained by applying the shooting method to Eq. \eqref{comp:52}.

\nocite{*}

\clearpage

\bibliography{bibitems}
\end{document}